\documentclass[12pt]{article}
\usepackage{amssymb}
\usepackage{epsfig,color}
\usepackage{pstricks,graphicx,epsfig,color,amssymb,amsmath,amscd}
\usepackage{cite}
\usepackage[]{graphicx}
\usepackage{makeidx}
\usepackage{amsmath}
\usepackage{nccmath}
\usepackage{setspace}

\usepackage[]{caption}
\renewcommand\rho{\varrho}
\newcommand{\bi}{\bibitem}
\newcommand{\be}{\begin{eqnarray}}
\newcommand{\ee}{\end{eqnarray}}

\begin{document}
\begin{titlepage}
\title{Antimatter in the Milky Way
}
\author{ A.\,D. Dolgov$^{a,b}$}

\maketitle
\begin{center}
$^a${Department of Physics, Novosibirsk State University,\\ 
Pirogova st.\,2, Novosibirsk, 630090 Russia}\\
$^b${{Bogolyubov Laboratory of Theoretical Physics, Joint Institute for Nuclear Research,\\
Joliot-Curie st.\,6, Dubna, Moscow region, 141980 Russia}}\\
\end{center}

\begin{abstract}

Recent astronomical observations indicating a strikingly abundant presence of antimatter in the Galaxy, 
in particular, of anti-stars are reviewed. Long-time earlier theoretical predictions are briefly discussed.

\end{abstract}
\thispagestyle{empty}
\end{titlepage}

\section{Introduction}

According to the conventional wisdom the universe in our neighborhood consists only of matter with 
a minor admixture of the secondary produced antiparticles in cosmic rays.
Anti-galaxies might exist, but they are far away from us. 

The first option is based on the Sakharov mechanism of baryogenesis~\cite{ADS-BG}, which in the
simplest version leads to a constant value of the baryon asymmetry over all the universe:
\be
 \beta = n_B/n_\gamma \approx 6\cdot 10^{-10} = const .
 \label{beta}
 \ee
 where $n_\gamma = 410.7/$cm$^3$ is the number density of CMB photons and
 $n_B = (n_b - n_{\bar b})$ is the difference between the number density of baryons and antibaryons.
It is believed that in the contemporary universe  $n_{\bar b} = 0$, while in the very early universe 
the density of baryons and antibaryons were almost the same:
\be
n_b  =n_{\bar b} + \delta n 
\label{n-delta-n}
\ee
where $\delta n \ll n_b$.
Slight excess of baryons over anti-baryons, $\delta n $,
 survived annihilation and made the building blocks of the present day universe.
  
Depending upon possible mechanisms of C and CP violation in cosmology~\cite{AD-CP}, another
possibility could be realised, namely that the universe consists of cosmologically huge domains of matter
and antimatter (galaxies and anti-galaxies) far away from each other.

In either case there would be galaxies and anti-galaxies consisting respectively
purely either of matter or antimatter
 but not of both. Observations of several recent years seem to create a strong doubt  about this statement,
since they lead to the conclusion that our Galaxy may be rich of antimatter. These data confirm our old
prediction~\cite{AD-JS,DKK} that anti-stars may populate the Milky Way. The mechanism suggested
in refs~\cite{AD-JS,DKK} leads to formation of primordial black holes (PBH) with the properties which
very well agree with the data. This agreement serves as justification of an accompanying
mechanism of the prediction of antistars in "our back yard".

\subsection{Anti-observations \label{s-obs}}

There are several seemingly independent pieces of astronomical data which unambiguously prove that there are 
plenty of antimatter in Milky Way. Quite intensive flux of gamma rays with energy  0.511 MeV speaks
of abundant population of positrons. Formerly the origin of positron was explained by $e^+e^-$-pair creation in 
strong magnetic fields of pulsars. However, the data AMS certainly exclude such an explanation. AMS group also 
reports on surprisingly high flux of antihelium nuclei. The last but not the least is the announcement of 
possible discovery of 14 antistars in the Galaxy~\cite{antistars}. Highly surprising and not expected 
by the community, this phenomena has
been predicted ages ago in our papers~\cite{AD-JS,DKK}.

\subsubsection{Cosmic positrons and antiprotons \label{ss-positrons}}

The flux of 0.511 MeV photons has been observed for 
pretty long time, staring from 1972~\cite{511-1}. Over the past 40
years there have been numerous publications on the observation of this line, 
see e.g.~\cite{511-2} and \cite{511-3}
and references therein for earlier measurements. 
Most precise data are presented by SPI/INTEGRAL~\cite{511-4,511-5,511-6}. 
According to these measurements the gamma ray flux 
from the central region of the Galaxy is
\be
\label{flux} 
\Phi_{511 \; {\rm keV}} = 1.07 \pm 0.03 \cdot 10^{-3} \; 
{\rm photons \; cm^{-2} \, s^{-1}}
\ee
with the width of about 3 keV.  This demonstrates that the electron--positron annihilation proceeds
at a surprisingly high rate.
The 0.511 MeV line emission is significantly detected towards the galactic bulge region 
and, at much lower level, from the galactic disk.

The registration of the 0.511 MeV line is a clear indication of electron-positron annihilation at rest.

Until recently the accepted hypothesis suggested for the explanation of the  abundant positron population 
in the Galaxy was that
$e^+$ were created in strong magnetic fields of pulsars but the observation of antiproton and 
positron energy spectra
made by AMS seems to exclude this possibility.
As it was announced by S. Ting  at  L'Aquila Joint Astroparticle Colloquium, 10th November, 2021 that the 
high energy spectra, $E >100$ GeV, of antiprotons and positrons are identical. It implies the same mechanism 
of $e^+$ and $\bar p$ origin. Since antiprotons cannot be created by pulsars, it means that neither can be created 
the positrons. This is a strong argument in favour of primordial antimatter 

\subsubsection{Anti-nuclei \label{ss-anti-nucl}}

In 2018 AMS-02 has also reported possible registration of six anti-helium-3 and two anti-helium-4 events:
A. Choutko,  AMS-02 Collaboration, “AMS Days at La Palma, La Palma, Canary Islands, Spain,” (2018) and
S. Ting," Latest Results from the AMS Experiment on the International Space Station", Colloquium at CERN, May, 2018.
The new anti-helium events are reported by S. Ting at
L'Aquila Joint Astroparticle Colloquium at 10th November, 2021.
According to his statement the fraction of anti-helium  at high energies was 
estimated as ${\bar He/He \sim10^{-9}}$, which is extremely high. 
It is not excluded that the total flux of anti-helium may be much higher because low energy anti-nuclei 
could avoid registration in the detector.

The secondary production of ${\bar{He}^4}$ is negligible.
The probability of the secondary production of light anti-nuclei was estimated in ref.~\cite{anti-nucl}.
Anti-deuterium could be produced in ${\bar p\, p}$
or ${\bar p\, He}$ collisions. The predicted flux of anti-deuterium is
${\sim 10^{-7} /m^{2}/ s^{-1} /sr/(GeV/n)}$,
i.e. 5 orders of magnitude lower than the observed flux of antiprotons.
The expected fluxes of secondary produced 
${^3\bar He}$ and ${^4\bar He}$ 
are respectively 4 and 8 orders of magnitude smaller than the flux of anti-deuterium.

\subsubsection{Anti-stars \label{ss-anti-nucl}}

This year a possible discovery of anti-stars in the Galaxy was reported~\cite{anti-star-disc}.
Quoting the authors:
“We identify in the catalog 14 antistar candidates not associated with any objects belonging to established 
gamma-ray source classes and with a spectrum compatible with baryon-antibaryon annihilation”
with characteristic energies of several hundred GeV. 
This sensational statement nicely fits the prediction of refs.~\cite{AD-JS,DKK}.

In ref.~\cite{bbbdp} a complementary method of antistar identification in the Galaxy was suggested. Namely it
was noticed that formation of atomic-like systems, similar to positronium, such as 
${p\bar p}$, ${p \bar{He}}$, ${He \bar{He}}$, etc preceded the annihilation. These "atoms" was 
formed in highly excited states and in the process of de-excitation they
emitted X-rays with the energies in the 1-10 keV range. These narrow lines can be associated with the hadronic 
annihilation gamma-ray emission. The most significant are L (3p-2p) 1.73 keV line (yield more 
than 90\%) from ${p \bar p}$ atoms, and M (4-3) 4.86 keV (yield $\sim 60$\%) and L (3-2) 11.13 
keV (yield about 25\%) lines from $He^4 \bar p$ atoms. These lines can be probed in dedicated 
observations by forthcoming sensitive X-ray spectroscopic missions XRISM and Athena and in 
wide-field X-ray surveys like SRG/eROSITA all-sky survey.

The bounds on the antistar density in the galaxy were studied in refs.~\cite{bound-1,bound-2,bound-3}
and happened to be not particularly restrictive because the annihilation could proceed 
only on the surface of anti-stars (or close to it in stellar wind)
due to a small mean free path of protons or nuclei, in contrast to disperse 
clouds of antimatter, where the annihilation can proceed in whole volume. That's why these anti-clouds,
which may also be created in the early universe, did not survive to our time, but disappeared much earlier.

The bound on the antistar density in solar neighbourhood was derived in ref.~\cite{von-bal}. 
Analysing the gamma ray luminosity induced by
Bondi accretion of interstellar gas to the surface of an antistar 
put  a limit on the relative density of antistars equal to
{${N_{\bar *} / N_{*} <  4\cdot 10^{-5} }$} inside 150 pc from the Sun. Still there is enough room for antistars in 
the Milky way (and presumably in other galaxies), so there are good chances to catch them.

\section{PBH observation \label{s-pbh}}

The next issue is how much can we trust theoretical model~\cite{AD-JS,DKK} which predicted antistars  in the Galaxy.
Antistars appeared as a by-product of this model where a new mechanism of creation of very massive 
primordial black holes (PBH) was proposed. A new feature of the suggestion is that it heavily relies on 
cosmological inflation.
Slightly later inflationary input for PBH formation was also invoked in refs~\cite{infl-1,infl-2,infl-3}.

Since the properties of PBH produced by the mechanism of refs.~\cite{AD-JS,DKK} are tested by observations and
happened to well agree with them, one may hope that another feature of the mechanism, namely existence of
anti-stars in the  Galaxy, has a good chance to be true as well.

The mechanism of refs.~\cite{AD-JS,DKK} predicts the log-normal mass spectrum of PBH, the only one
among several calculated spectra of PBH,  
which is verified and strongly supported by the data. The number density of PBH as a function of 
mass, $M$, is given by the function: 
\be
\frac{dN}{dM} = \mu^2 \exp{[-\gamma \ln^2 (M/M_0)]. }
\label{log-norm}
\ee
It depends upon three constant parameters, out of which one is theoretically predicted:
$M_0 \approx 10 M_\odot$~\cite{AD-KAP-M0}. 

In ref.~\cite{chirp}  the available data on the chirp mass distribution of the black holes in the coalescing binaries in 
O1-O3 LIGO/Virgo runs are analysed and compared with theoretical expectations based on the hypothesis that these black holes are primordial with  log-normal mass spectrum. The results are presented in Fig. 1

\begin{figure}[htbp]
\begin{center}
\includegraphics[scale=0.2]{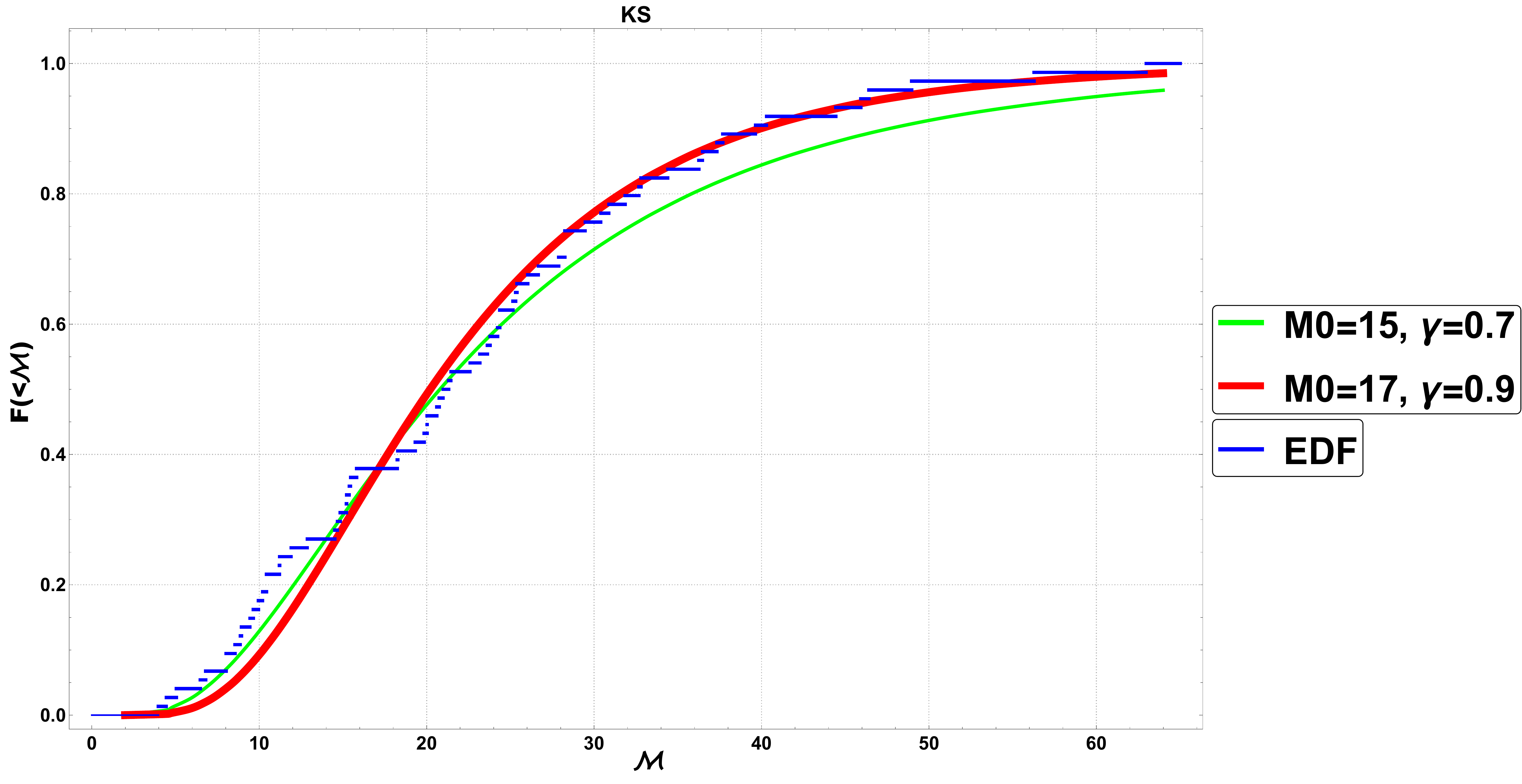}
\caption{Model distribution $F_{PBH}(< M)$ with parameters  $M_0$ and ${\gamma}$ for two best 
Kolmogorov-Smirnov tests.  EDF= empirical distribution function.}
\end{center}
\end{figure}

{The inferred best-fit mass spectrum parameters, {$M_0=17 M_\odot$ 
and ${\gamma=0.9}$,} fall 
within the theoretically expected range and shows excellent agreement with observations.} 
{On the opposite, binary black hole 
models based on massive binary star evolution require additional adjustments to reproduce the observed 
chirp mass distribution, as it is seen from Fig. 2

\begin{figure}[htbp]
\begin{center}
\includegraphics[scale=0.25]{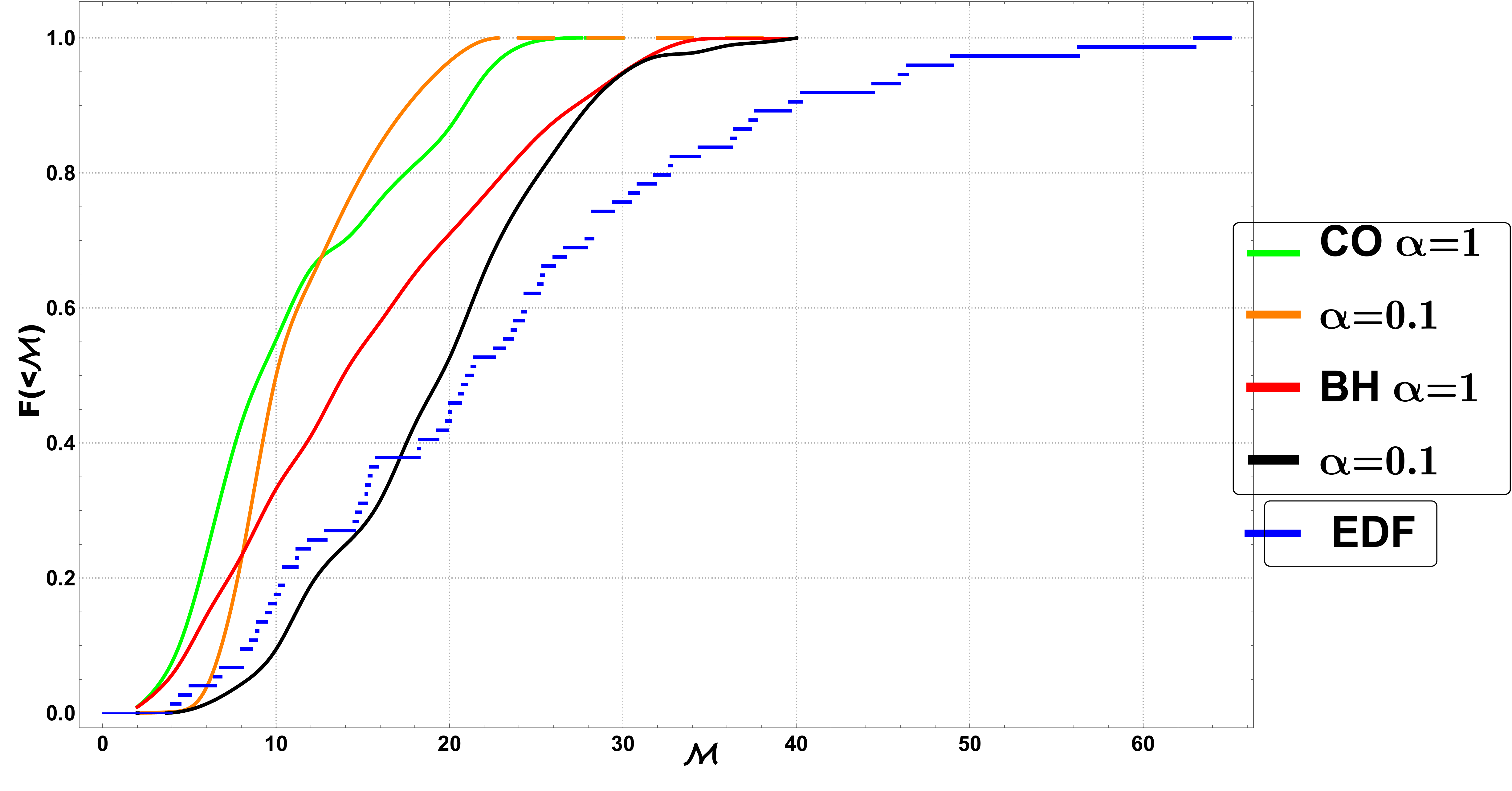}
\caption{Cumulative distributions $F(< M)$ for several astrophysical models of binary BH coalescences.}
\end{center}
\end{figure}


Hence we can conclude that 
 PBHs with log-normal mass spectrum perfectly fit the data, while
astrophysical BHs seem to be in tension with the observed events..

An additional argument in favour of PBH formation by the mechanism of ref.~\cite{AD-JS,DKK}
with log-normal mass spectrum
comes from the data on the observation of black holes with intermediate masses with
$M = (10^2 - 10^5)  M_\odot$~\cite{AD-SP}.
A review of the conventional cosmological and astrophysical problems, which are uniquely settled 
down by the assumption that black holes in the universe are mostly primordial, can be
found in ref.~\cite{AD-UFN}.

 \section{A few words about the mechanism \label{theor}}
 
Let us briefly describe here the basic features of the mechanism of antimatter
creation~\cite{AD-JS,DKK}. More detail can be found
in the original papers. The model is based on the well
known Affleck-Dine baryogenesis scenario~\cite{Affl-Dine}  which utilises an assumption of high energy scale 
supersymmetry. In such a theory
there exists scalar field $\chi$ with nonzero baryonic number.
The potential of $U(\chi)$ generically have the so called flat directions along which $U$ does not rise, and so
$\chi$ may reach quite large values leading to a very high baryon asymmetry.
Our new input  to this mechanism is an assumption of the
interaction between $\chi$ and the inflaton field $\Phi$ of the form
\be
L_{int} = \lambda |\chi|^2 \left(\Phi - \Phi_1\right)^2,
\label{chi-Phi-int}
\ee
where $\Phi_1$ is the value of the inflaton field reached in the course of inflation. Due to this interaction
the gates to the flat directions happened to be closed, except for a brief period when $ \Phi \approx \Phi_1$ 
and so the 
possibility for $\chi$ to reach a large value  became open. When the gates were closed $\chi$ lived
near $\chi = 0$, but  when the gates became open, $\chi$ could reach a large value during this "happy" time, but
it had to return back to zero when $\Phi$ overpassed $\Phi_1$. Due to misalignment of quadratic and quartic
flat direction $\chi$ started to "rotate" in complex $\chi$-plane and it means that it might acquire a huge
 baryonic number in a bubble of cosmologically small size but possibly astrophysically large. 
 Thus we come to relatively small bubbles with high density of baryons or antibaryons..

The emerged universe looks as Swiss cheese "upside down". Almost all volume is devoid of baryons or to be more 
precise the average baryonic number in the bulk of the universe is very small, 
$\beta \equiv n_B/n_\gamma \approx 10^{-9}$, while there exists dense bubbles with high $\beta \sim 1$. 

In the very, early universe the density contrast between the bubble and the bulk was practically 
zero due to vanishing quark mass.
However the density contrast dramatically rose up after the QCD phase transition when massless (or very light)
quarks turned into heavy nonrelativistic baryons. Hence after the QCD phase transition the bubbles with high density
of baryons turned mostly into into black holes or, if not enough massive, 
a fraction of them became compact stars or antistars, depending upon
the direction of rotation of $\chi$ in the complex $\chi$-plane. 

\section{Conclusion \label{concl}}

Surprisingly large amount of antimatter in our Galaxy demonstrated by the 0.511 MeV gamma rays
from $e^+e^-$-annihilation, anti-helium in high energy cosmic rays, and possible observation or 14 
anti-stars present very strong evidence in favour of the theoretical predictions made in 
refs.~\cite{AD-JS,DKK}.

More tests and data are of course necessary. In particular, an alternative search for anti-stars  
through narrow 1-10 keV X-rays lines, proposed in ref.~\cite{bbbdp}, can certainly give valuable addition
to the existing methods of anti-star registration.

The mechanism~\cite{AD-JS,DKK} predicts creation of several kind of anomalous stars, having unusual
chemistry, too old, or too fast. There is a good chance that these stars may consist of antimatter.

\section*{Acknowledgement}
The work was supported by the  by RSF Grant 20-42-09010.

\end{document}